# Detection of Replay Attacks to GNSS based on Partial Correlations and Authentication Data Unpredictability

Gonzalo Seco-Granados, David Gómez-Casco, José A. López-Salcedo

IEEC-CERES, Universitat Autonoma de Barcelona (UAB), Spain

Email: {gonzalo.seco, david.gomez.casco, jose.salcedo}@uab.cat

Ignacio Fernández-Hernández, European Commission, DG GROW, Belgium.

Email: ignacio.fernandez-hernandez@ec.europa.eu

**Abstract**

Intentional interference, and in particular GNSS spoofing, is currently one of the most significant concerns of the Positioning, Navigation and Timing (PNT) community. With the adoption of Open Service Navigation Message Authentication (OSNMA) in Galileo, the E1B signal component will continuously broadcast unpredictable cryptographic data. This allows GNSS receivers not only to ensure the authenticity of data origin but also to detect replay spoofing attacks for receivers already tracking real signals with relatively good visibility conditions. Since the spoofer needs to estimate the unpredictable bits introduced by OSNMA with almost zero delay in order to perform a Security Code Estimation and Replay (SCER) attack, the spoofer unavoidably introduces a slight distortion into the signal, which can be the basis of a spoofing detector. In this work, we propose five detectors based on partial correlations of GNSS signals obtained over predictable and unpredictable parts of the signals. We evaluate them in a wide set of test cases, including different types of receiver and spoofing conditions. The results show that one of the detectors is consistently superior to the others, and it is able to detect SCER attacks with a high probability even in favorable conditions for the spoofer. Finally, we discuss some practical considerations for implementing the proposed detector in receivers, in particular when the Galileo OSNMA message structure is used.

**Introduction**

Global Navigation Satellite System (GNSS) spoofing attacks are intentional interference, whose aim is to manipulate the Position, Velocity and Time (PVT) of a target GNSS receiver. Data authentication, such as that provided by Galileo Open Service Navigation Message Authentication (OSNMA) in the E1B component, allows GNSS receivers to detect data spoofing attacks. If the value of the authenticated bits is not correct, the receiver realizes that the received signal is not authentic. However, spoofers can still forge the position by replaying the signals and thus altering the range measurements. To protect against range alteration, one can exploit the fact that a large percentage of OSNMA cryptographic information is unpredictable. Assuming that the receiver can regenerate and verify the unpredictable symbols encoding OSNMA bits, this unpredictability greatly increases the complexity of possible range spoofing attacks (Fernández-Hernández and Seco-Granados 2016). However, some attacks may still be possible if the signal is not carefully processed. In fact, Security Code Estimation and Replay (SCER) attacks can be applied (Humphreys 2013). These attacks consist of two steps. First, the spoofer tracks the received signals from the satellites and estimates the symbols transmitted by the satellites in view, including unpredictable symbols. Second, the spoofer reradiates GNSS-like signals based on these estimations to the target GNSS receiver in order to take control of the tracking loops, and eventually, the user position. After these spoofing attacks were proposed, different views have emerged on the degree with which data unpredictability can provide range protection. In particular, there has been some controversy about the potential usefulness of the unpredictability derived from the authentication of the message. On the one hand, Psiaki and Humphreys (2016) considered that Navigation Message Authentication (NMA) together with SCER detection based on a careful processing of the NMA bit at the receiver may have a similar efficiency against spoofing as code-based authentication. On the other hand, Caparra et al. (2017) or Curran and O'Driscoll (2016) raised doubts on the effectiveness of NMA to protect range measurements.

Generating a SCER attack is far from trivial for the spoofer since the spoofed signal must be synchronized with the authentic signal. If the two signals are not aligned with each other in the time domain when the spoofer starts the attack, the receiver will detect a clock jump. This occurs because the stability of the receiver clock can be known, and hence high variations of clock offset in a short period of time at the PVT stage may be caused by a spoofer. More details on clock stability can be found in Beard and Senior (2017). In order to perform the SCER attack without forcing a clock jump, or blocking the true signal in the receiver for sufficient time to estimate it, the spoofer needs to perform a zero (or almost zero) delay SCER attack, which is based on transmitting a signal that is practically synchronized with the true signal. This implies that, prior to the implementation of SCER protection, receivers should perform simple checks on the signal power, including jamming detection, and consistency of measurements and PVT. In any case, replay attacks are not a marginal threat, but they are probably the main vulnerability of receivers not processing encrypted signals; and nowadays, encrypted GNSS signals for civilian use do not exist. Furthermore, the particular case of zero (or almost zero) delay replay attacks is the case worth being studied because if the spoofer introduces a noticeable delay, then it can be detected through other means (Psiaki and Humphreys 2016).

Even so, a spoofer succeeding in a zero-delay SCER attack may take control of the receiver tracking loops, and therefore of the receiver position. Nevertheless, by definition, the spoofer cannot a priori know the value of the unpredictable symbols, and hence the synchronized-yet-spoofed signal will include some errors during the first microseconds of the unpredictable symbols, that is to say, during the period of time while the spoofer has not accumulated enough energy to have a reliable estimation of the new symbol. The questions on whether these errors permit the GNSS receiver to detect the spoofing attack, and how many unpredictable symbols are needed to have a reliable detector, remain still open, but we will try to clarify the issue significantly.

The main objectives of this work are to define new antispoofing techniques based on unpredictable symbols and second, and next to assess the effectiveness of these techniques in a more exhaustive way than that shown in the current literature.

First, we propose a signal model that includes the real and spoofed signals. We also introduce and justify the spoofing attack model used in our simulations. Second, we present five spoofing detectors based on the comparison of the received samples in the unpredictable and predictable parts of the signal. We also present the motivation and expected advantages and disadvantages of each detector. Third, we evaluate the performance of the detectors under the proposed spoofing attack and several conditions, including different power levels, channels, and correlation times. We identify the best-performing detector and its sensitivity to different situations.

Fourth, we look at the practical implementation and constraints that the user receiver and the spoofer would experience, and the adequacy and robustness of the proposed antispoofing techniques in real cases. In particular, we focus on determining the time required by the target receiver to obtain enough unpredictable symbols to detect spoofing attacks based on the currently proposed Galileo OSNMA protocol. We also analyze the time a spoofer needs to have a reliable estimation of the symbol, to delay the spoofed signal with respect to the authentic signal without being noticed by comparison with the receiver clock. Finally, we present the conclusions of the work. Our results show that, if a receiver is already tracking real GNSS signals, symbol unpredictability can be a robust measure against spoofing attacks based on signal replay.

**Signal and attack model**

Spoofing detection is treated as a binary hypothesis testing problem. It can be modeled under the hypotheses that the spoofer is present ($H_1$) or absent ($H_0$) as follows. The received sampled signal for one satellite can be expressed as:

$$y(n) = \begin{cases} Ab(n)c(n)e^{j(2\pi f_d^r n + \varphi_d^r)} + \beta \tilde{b}(n)c(n)e^{j(2\pi f_d^s n + \varphi_d^s)} + \omega(n), & (H_1) \\ Ab(n)c(n)e^{j(2\pi f_d^r n + \varphi_d^r)} + \omega(n), & (H_0) \end{cases} \quad (1)$$

The real (that is, authentic or legitimate) signal amplitude is $A$, $\beta$ is the amplitude of the spoofing signal, $b(n)$ is the unpredictable symbol, $c(n)$ is the pseudorandom noise code, $f_d^r$ is the Doppler frequency and $\varphi_d^r$ the phase of the real signal for satellite, $\tilde{b}(n)$ is the unpredictable symbol transmitted by the spoofer, and $\omega(n)$ is additive white Gaussian noise. It is clear that the total received signal should include the sum for all visible satellites and spoofed signals, but for the sake of clarity, we have written only one in the formulation. We consider that the spoofer is performing a zero-delay attack and therefore the spoofer introduces no signal delay and transmits at the same frequency for a given satellite, so $f_d^r = f_d^s$. The amplitude $A_p$ and the phase $\varphi_d^r$ can be different from $\beta_l$ and $\varphi_{d,l}^s$. We assume that our spoofer can control the spoofed signal amplitude $\beta$ and make it in some cases equal to $A$, as will be shown later, but it cannot align the carrier phase measurement to the real one, as aligning carrier phase measurements would require an extremely high level of accuracy, if feasible at all.

The notation that we will use in the simulations is illustrated in Fig. 1. This figure shows a representative example of a spoofing attack for one satellite. The definition of each parameter can be found in Table 1. Note that, in this work, we focus on a single spoofing signal for only one satellite. If the proposed spoofing metric is effective for detecting one spoofing signal, it will be even more difficult for the spoofer to consistently spoof a full PVT solution, as this would require succeeding for multiple satellite signals at the same time.

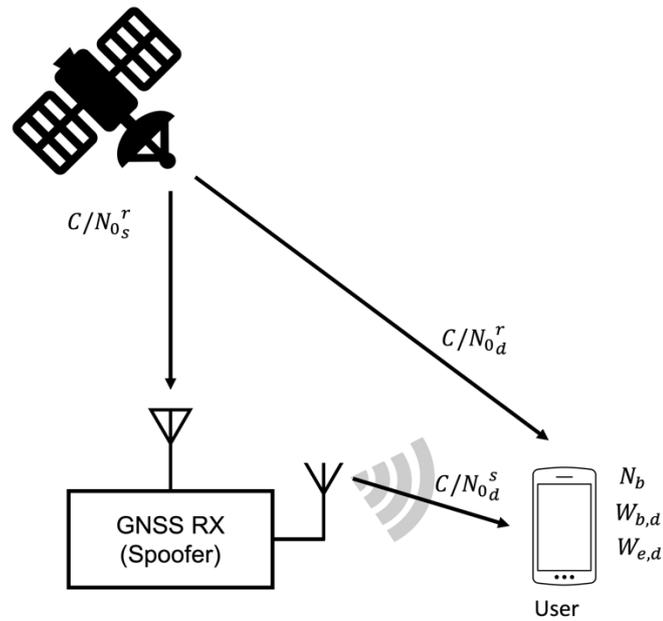

**Fig. 1** Illustration of a spoofing attack showing the main variables that characterize the spoofer and the detector

**Table 1** Spoofing zero-delay attack on GNSS signals with unpredictable symbols - Parameter definition

| Variables | Definition |
|---|---|
| $C/N_{0_s}^{r}$ | $C/N_0$ of the real signal received by the spoofer. |
| $C/N_{0_d}^{r}$ | $C/N_0$ of the real signal received by the user or detector |
| $C/N_{0_d}^{s}$ | $C/N_0$ of the spoofed signal received by the user or detector |
| $N_b$ | Number of unpredictable symbols used in the spoofing detection techniques. Note that we use the term "unpredictable symbol" to refer to the portions of the amplitude modulation of the received |

|  | signal which are unpredictable, not to the unpredictable data that generates the symbols after a coding process. |
|---|---|
| $W_{b,d}$ | Duration of the partial cross-correlation used at the beginning of the symbol. |
| $W_{e,d}$ | Duration of the partial cross-correlation used at the end of the symbol. |

Before describing the implementation of zero-delay attacks in more detail, we describe here our attack model assumptions. First, we assume that the receiver is tracking authentic signals at the start of the attack, i.e., the receiver starts up and performs acquisition in a controlled environment. Spoofing a receiver at the acquisition phase is out of the scope of this work. Many other spoofing analyses, such as Humphreys (2013) or Hegarty et al. (2019), also focus on this use case, where it is assumed that the startup has been performed safely. Second, we assume that, in the current attack, the spoofer does not force signal reacquisition. The reason is that, as it will be shown in later sections, the spoofer would have to prevent the target receiver from tracking the authentic signal during several minutes. This means that the spoofer would have to jam the receiver, which would enter into reacquisition stage, and to ensure that the receiver does not resume normal operation in less than several minutes. This period of time would cause enough time uncertainty in the target receiver so that the spoofer can delay the signal, estimate the unpredictable symbols, and thus take control of the target receiver tracking loops without being uncovered. The computation of the interval of minutes assumes that a standard TCXO (temperature-controlled crystal oscillator) is used in the receiver. The length of the interval is increased in orders of magnitude if an oven-controlled crystal oscillator (OCXO) is used. Therefore, our attack model focuses on a spoofed signal aligned with the real signals in code delay and frequency, in order to take control of the tracking loops without forcing reacquisition. Note that, in these conditions, taking control of the loops would lead to cycle slips, but analyzing cycle slips as part of the detector is left for further work.

The weakness of zero-delay attacks is that the signal transmitted by the spoofer unavoidably includes some errors in the first part of the unpredictable symbols. In order not to be detected easily by the target receiver, the spoofer can replay the signal in three different ways:

- Estimated value: Trying to estimate the unpredictable symbol sample by sample and introducing this estimation in the spoofed signal. By doing so, the first part of the symbol would contain several changes of the sign because it is not feasible to obtain a reliable estimation of the symbol, but after a certain number of samples, the spoofer generates the real value of the unpredictable symbol with very large likelihood. An example of this attack is shown in the top panel of Fig. 2.

- Random value: Introducing a random value of 1 or -1 at the beginning of the symbol during a short period of time, and once the spoofer has a reliable estimation of the unpredictable symbol value, it is included in the rest of the symbol (Fig. 2 bottom).

- Zero value: Introducing a value of zero at the beginning of the symbol during a short period of time, and once the spoofer has estimated the unpredictable symbol value, it is included in the rest of the symbol. Note that, during the zero-value time, the spoofed signal is not transmitted (Fig. 2 bottom).

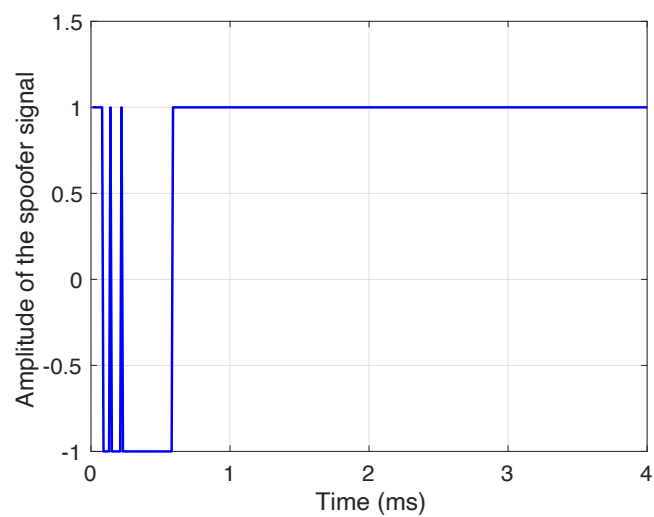

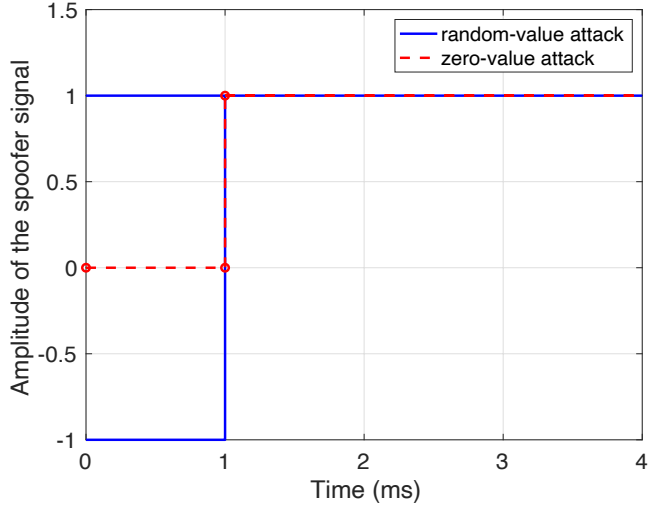

**Fig. 2** Amplitude of the spoofer signal during one unpredictable symbol: the estimated-value attack in the top panel, random-value and zero-value attacks in the bottom panel.

**Spoofing detection techniques**

The proposed spoofing detection techniques deal with the comparison between the initial part of an unpredictable symbol and other parts of the signal that are considered predictable, such as predictable symbols or the last part of unpredictable symbols. The Galileo E1B signal uses a BPSK-like modulation, although it is modulated with a 1 MHz sub-carrier in order to generate the Binary Offset Carrier BOC(1,1) signal (European Union 2016). Each 4-millisecond code is modulated with a symbol at 250 symbols/s, which transmits the navigation message of 120 bits/s, plus 10 synchronization symbols per second. In the rest of this work, we will refer to bits or symbols, depending on the context, but for the specific case of Galileo E1B we refer to the E1B data *symbols*, not to the bits resulting from the convolutional decoding, but the techniques can be applied without modification to the bits of other unencoded messages. The core idea of the techniques, as already pointed out in Humphreys (2013) and Fernández-Hernández and Seco-Granados (2016), is that the spoofer will have difficulties to estimate the unpredictable part of the signal, and therefore, the correlation results will be worse than those on the predictable part. The techniques used herein are based on comparing the initial part of the unpredictable symbol (unpredictable part) with the last part of the unpredictable symbol (predictable part), although similar results can be obtained if the predictable part is taken from predictable symbols, in order to introduce some randomness. This randomness makes useless any attempt of the attacker to degrade other parts of the signal to mimic the degradation introduced in the unpredictable parts. As the proposed methods are applied directly on the received signal, attacks related to the decoding process, as those proposed in Curran and O'Driscoll (2016), are not effective in our case.

The spoofing detector in the receiver computes the partial correlations at the beginning and the end of an unpredictable symbol, respectively, which can be expressed as

$$B'_{beg}(k) = \sum_{n=1}^{N} y_{beg,k}(n) x^*_{beg}(n) \qquad (2)$$

$$B'_{end}(k) = \sum_{n=1}^{N'} y_{end,k}(n) x^*_{end}(n) \qquad (3)$$

where $y_{beg}(n)$ and $y_{end}(n)$ are the first and the last samples, during intervals of time $W_{b,d}$ and $W_{e,d}$ (or $N$ and $N'$ in numbers of samples), respectively, of the received signal in the $k$-th unpredictable symbol, and $x^*_{beg}(n)$ and $x^*_{end}(n)$ are the corresponding local replicas, where (*) denotes the complex conjugate. If the zero-delay spoofer is present, some differences can be found between the initial and the late partial correlations. We define $B_{beg}(k)$ and $B_{end}(k)$ as the partial cross-correlation after removing the sign of the unpredictable symbol, $b(k) \in \{-1,1\}$:

$$B_{beg}(k) = b(k) B'_{beg}(k) \qquad (4)$$

$$B_{end}(k) = b(k) B'_{end}(k) \qquad (5)$$

Note that the estimation of the symbol can be assumed to be free of error because the detector can be applied once the user receiver has checked that the message is cryptographically correct. The detectors are described hereafter. In all cases, the objective is to propose and analyze via simulations the potential use as spoofing detectors of simple combinations of values that are readily available at the receiver. The derivation of the statistical hypotheses of the detection problems that would result in each of the detectors is beyond the scope of the paper, but it is an interesting line for future work. An intuitive way of detecting spoofing would be to compare the dispreading gain based on several unpredictable parts with that obtained from various predictable parts. One way of performing this comparison is computing, first, the ratio of partial correlations accumulated over $N_b$ symbols. Then, the absolute value of the ratio between the two metrics is computed as

$$R_1 = \left| \frac{\sum_{k=1}^{N_b} B_{beg}(k)}{\sum_{k=1}^{N_b} B_{end}(k)} \right| \qquad (6)$$

If only the spoofing signal is received, the value of $B_{beg}(k)$ tends to be small, and therefore $R_1$ also tends to be small. However, a drawback of the $R_1$ metric is that it can provide any value in $H_1$ if the received signal includes the spoofed signal and the real one with different phases. As the real and spoofed signals can be combined constructively or destructively depending on their relative phases, $B_{end}(k)$ can take both small and large values in the presence of spoofing, resulting in a wide range of possible values for $R_1$, larger and smaller than 1, overlapping with the value that $R_1$ takes in the absence of spoofing, that is, $R_1 \approx 1$. This complicates the definition of the detection threshold since, in principle, detecting that $R_1$ is small is not sufficient to identify all spoofing events, inherently causing a large probability of misdetection. This effect is illustrated in Fig. 3, which depicts that the value of $B_{end}(k)$ results from the addition of the complex partial correlations contributed by the authentic signal and by the spoofer. As the phase difference $\Delta\varphi$ can be arbitrary, the value of $B_{end}(k)$ can change also arbitrarily with respect to one obtained when only the authentic signal is present. The same discussion is valid for $B_{beg}(k)$.

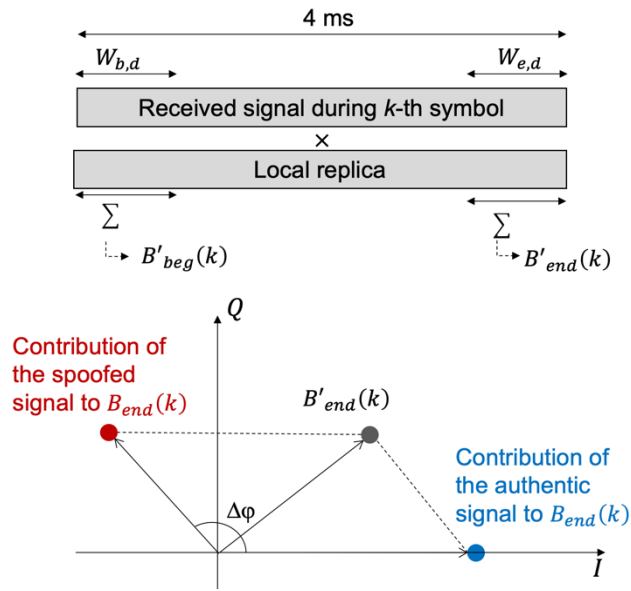

**Fig. 3** Example of the effect of the combination of real and spoofed signals in the correlator output, for a phase difference between the real and spoofed signal of $\Delta\varphi$.

In order to solve this problem, we propose another alternative as follows:

$$R_2 = \left| \frac{\sum_{k=1}^{N_b} B_{beg}(k)}{\sum_{k=1}^{N_b} B_{end}(k)} - 1 \right| \qquad (7)$$

The idea behind $R_2$ is that, if the spoofer is absent, $R_2$ is close to 0, but if the spoofer is present, $R_2$ becomes larger. This facilitates the definition of the detection threshold, avoiding the overlap between values of both hypothesis that occurs with $R_1$.

An additional method is $R_3$, which consists in computing the mean of the difference between the initial and final partial correlations. If $R_3$ is large, the spoofer is present, and if $R_3$ is small, the spoofer is absent.

$$R_3 = \left| \frac{1}{N_b} \sum_{k=1}^{N_b} (B_{beg}(k) - B_{end}(k)) \right| \tag{8}$$

Note that $R_2$ is equivalent to $R_3$, but without dividing by $\frac{1}{N_b} \sum_{k=1}^{N_b} B_{end}(k)$. As we will show, this improves the detection capability as $R_3$ is less sensitive to the predictable part of the symbol. Detector $R_1$ did not depend on the phases of the terms $\sum_{k=1}^{N_b} B_{beg}(k)$ or $\sum_{k=1}^{N_b} B_{end}(k)$, only on the relative amplitudes. On the contrary, detectors $R_2$ and $R_3$ make use of both the amplitudes and the phases of those partial correlations, and this is an indication that $R_2$ and $R_3$ are potentially more informative about the presence or absence of spoofing.

Another interesting option consists in comparing the carrier-to-noise (C/N$_0$) estimate obtained from the initial part of an unpredictable symbol to the estimate obtained from predictable parts. To estimate the C/N$_0$, we propose the well-known Narrow-band Wide-band Power Ratio (NWPR) estimator. It requires evaluating the ratio between the signal narrowband power to its wideband power (Van Direndonck 1996) (Gomez-Casco et al. 2018),

$$NP_x = \frac{NBP_x}{WBP_x} \tag{9}$$

where

$$WBP_x = \left( \sum_{k=1}^{N_b} |B_x(k)|^2 \right) \tag{10}$$

$$NBP_x = \left( \left| \sum_{k=1}^{N_b} B_x(k) \right|^2 \right) \tag{11}$$

where $B_x(k)$ is the partial correlation of any part of the symbol, i.e. $x \in \{beg, end\}$. Finally, the C/N$_0$ can be estimated as

$$\widehat{C/N_0}_x = 10 \log_{10} \left( \frac{1}{W_{x,d}} \frac{NP_x - 1}{N_b - NP_x} \right) \tag{12}$$

The detector compares the estimate of the first part of the unpredictable symbol ($\widehat{C/N_0}_{beg}$), which is considered to be unpredictable, to the estimate of the last part of the unpredictable symbol ($\widehat{C/N_0}_{end}$):

$$R_4 = \left| \widehat{C/N_0}_{beg} - \widehat{C/N_0}_{end} \right| \tag{13}$$

If the spoofer is absent, the metric above must be close to zero, while if present, it must provide larger values.

Finally, we present $R_5$, which is a detector that only uses the phases of the initial and final partial correlations. If the presence of the spoofed signal modifies the phase of the received signal between the unpredictable and predictable parts differently, the spoofer can be detected using the metric:

$$R_5 = |\psi_{beg}(k) - \psi_{end}(k)| \tag{14}$$

$$\psi_x(k) = atan2\left(\sum_{k=1}^{N_b} Im(B_x(k)), \sum_{k=1}^{N_b} Re(B_x(k))\right), \ x \in \{beg, end\} \tag{15}$$

where $atan2(\cdot,\cdot)$ is the four-quadrant arctangent, and $Re(\cdot)$ and $Im(\cdot)$ are the real and imaginary parts of the signal.

**Performance evaluation**

This section analyzes the capabilities of the five proposed techniques under the presence of zero-delay attacks and for the most relevant attack situations. One of the difficulties of the analysis is that the space of possible spoofing attacks and detector parameters configuration is very large. Therefore, the results presented correspond in principle to the most difficult-to-detect spoofing situations, in terms of spoofing power advantage and type of attack. The spoofing simulation parameters are presented in Table 2. Regarding the attack types, out of the three attacks previously described, we focus on the estimated-value attack, as it is the most sophisticated attack and the one that makes the spoofed signal closest to the real signal as early as possible. It, therefore, provides an upper bound for the required number of unpredictable symbols compared to the other two attacks. We recall that this attack consists in estimating the unpredictable symbol sample by sample and introducing this estimation in the spoofed signal. The estimation of the $k$-th unpredictable symbol carried out by the spoofer at the tracking stage can be expressed as:

$$\hat{b}_s(m;k) = \text{sign}\big(Re\{\textstyle\sum_{n=1}^{m} y_{beg,k}(n)x_{beg}^*(n)\}\big) \tag{16}$$

where $m$ represents the number of samples accumulated since the beginning of the symbol. By doing so, the spoofer obtains an estimation of the symbol for each successive sample $m$. A variant of this attack also consists in estimating the symbol sample by sample, but then transmitting the estimation of the symbol by using a scalar factor, depending on the level of confidence of the attacker (Humphreys 2013). This variant has been simulated as well, and the results have not significantly differed to the ones obtained with the described estimated-value attack.

We also assess the cases in which the spoofer has a *C/N₀* advantage of up to 5 dB with respect to the receiver. Concerning the relative power between the spoofed and real signals, as received in the receiver, we assess the cases of equal power and +3 dB for the spoofed signal. The results are tested for additive white Gaussian noise (AWGN) and land-mobile satellite (LMS) channels. In all case, the phase difference between the real and the spoofed signals is considered to be uniformly random in $[0,2\pi)$. Without loss of generality in the results, we use a moderate false alarm probability ($P_{fa}$ = 0.02) because it provides a reasonable baseline for comparing the detectors. The detection thresholds for each detector are determined experimentally to ensure that the real false alarm probability coincides with the target one. Even if a false alarm rate of 2% may seem too high for some applications and degrade availability if used in isolation, the proposed method can be combined with other indicators in order to filter out false alarms.

Table 2 Parameterization of spoofing simulations

| Zero-delay Attack type | "Est" |
|---|---|
| $C/N_{0_s}^{r}$ | 0 dB advantage; + 3 dB advantage; + 5 dB advantage with respect to $C/N_{0_d}^{r}$ |
| $C/N_{0_d}^{s}$ | Same power as $C/N_{0_d}^{r}$ ; +3 dB with respect to $C/N_{0_d}^{r}$ |
| $W_{b,d}$ ; $W_{e,d}$ | 0.125ms; 0.25ms; 0.5ms. |

| | | |
|---|---|---|
| Channel model | AWGN; LMS | |
| $P_{fa}$ | 0.02 | |

In all cases, the spoofing detection probability ($P_d$) is measured for different numbers of symbols $N_b$ under different combinations of these parameters. For the considered durations of the observation windows $W_{b,d}$ and $W_{e,d}$ the inter and intra-system interference proves to have a marginal effect, as it will be commented later on, compared to the additive noise, and hence only one Galileo E1B-E1C signal has been used in the simulation results.

*Results in an AWGN channel at different C/N₀ values*

Fig. 4 shows the probability of detecting the spoofing attack vs. the number of unpredictable symbols for $W_{b,d} = W_{e,d} = 250\mu s$. We consider that the user receives both the real and spoofed signals in the $H_1$ hypothesis, as in (1). The spoofer receives the real signal with $C/N_{0_S}^r$=40 dB-Hz, and the user receives both the real and spoofed signals with the same power, 40 dB-Hz (top plot) and 37 dB-Hz (bottom plot), that is, $\frac{C}{N_{0_d}^r} = \frac{C}{N_{0_d}^s} = \{37, 40\}$ dB-Hz. This means that the bottom plot corresponds to a case where the spoofer has a 3 dB advantage in signal reception with respect to the user. The top plot shows that the R2 and R3 techniques provide the best performance, and they require only about 125 and 110 symbols, respectively, to detect the spoofing attack with a probability of 0.9. On the other hand, when the spoofer has an advantage of 3 dB with respect to the user receiver, the R2 and R3 detectors need to observe approximately 100 unpredictable symbols more to detect the attack with the same probabilities of false alarm and detection, as seen on the bottom plot of Fig. 4. Detector R1, which would be the natural choice to detect differences between the partial correlations at the beginning and end of each unpredictable symbol, suffers a strong degradation in the typical case where the received signal contains the real and spoofed signal, as already anticipated. This occurs because, in the presence of both signals, the R1 metric can take practically any value depending on the phase difference between those signals, while in the absence of spoofing, the R1 metric takes values close to 1. This problem does not happen when the spoofed signal is significantly stronger than the real signal since, in this case, R1 takes values close to zero.

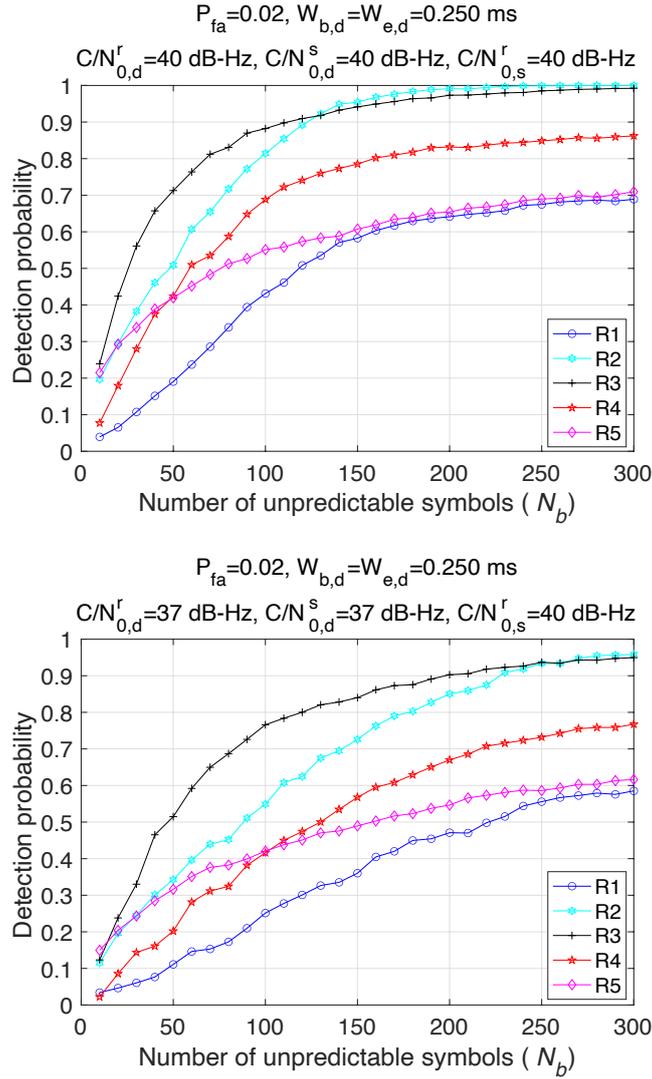

**Fig. 4** Detection probability vs the number of unpredictable symbols for false alarm probability of 0.02. On the top plot, the user and spoofer receive all signals at the same power. On the bottom plot, the spoofer has a 3 dB advantage.

The previous simulation considers that the user receives the signal from the spoofer and the satellite with the same power. In Fig. 5, we assume that the user receives the spoofed signal with 3 dB more than the real signal. In this test case, the user receiver can detect the spoofing attack more easily than in the previous simulation. In these conditions, the best detector is again R3. The number of symbols required by R3 to detect the spoofing attack with probability 0.9 is only 70, in contrast with the 200 symbols required for the case seen in the bottom plot of Fig. 4, where the spoofing signal had the same power as the real one. It is worth mentioning that in this particular test case, the performance of R1 is not so poor because, due to the power imbalance, the constructive and destructive combination between the spoofing and real signals is not so effective.

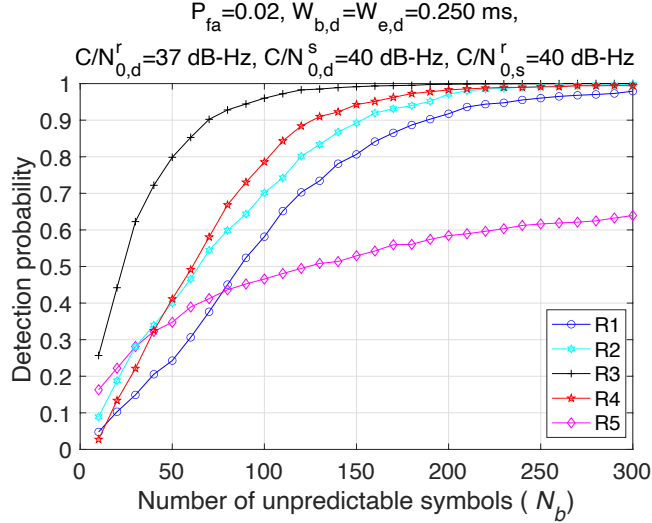

**Fig. 5** Detection probability vs. the number of unpredictable symbols for false alarm probability of 0.02. The spoofed signal is received at 3 dB more power than the real signal.

*Results in an AWGN channel for different lengths of the correlation interval*

In Fig. 6, we analyze how the performance of the detectors is affected by the use of different lengths of the windows used to compute the partial correlations: 0.125 ms (in Fig. 6-top), 0.250 ms (Fig. 4-bottom), and 0.500 ms (Fig. 6-bottom). These correlations, when accumulated for several symbols, while much shorter than the standard 4ms Galileo E1 codes, ensure that there is sufficient gain for detection, even in case of cross-correlation interference from other satellites is considered. The results show that R3 provides very similar performance for different window lengths used to compute the partial correlations, while the others are more sensitive to this parameter. This has some practical implications that are discussed in the next section. Detector R2, which also exhibited promising performance, is more affected by the window length. Only if the window length is appropriate, it offers good performance. However, when the time window is too short or large, this technique suffers a clear degradation in the detection probability.

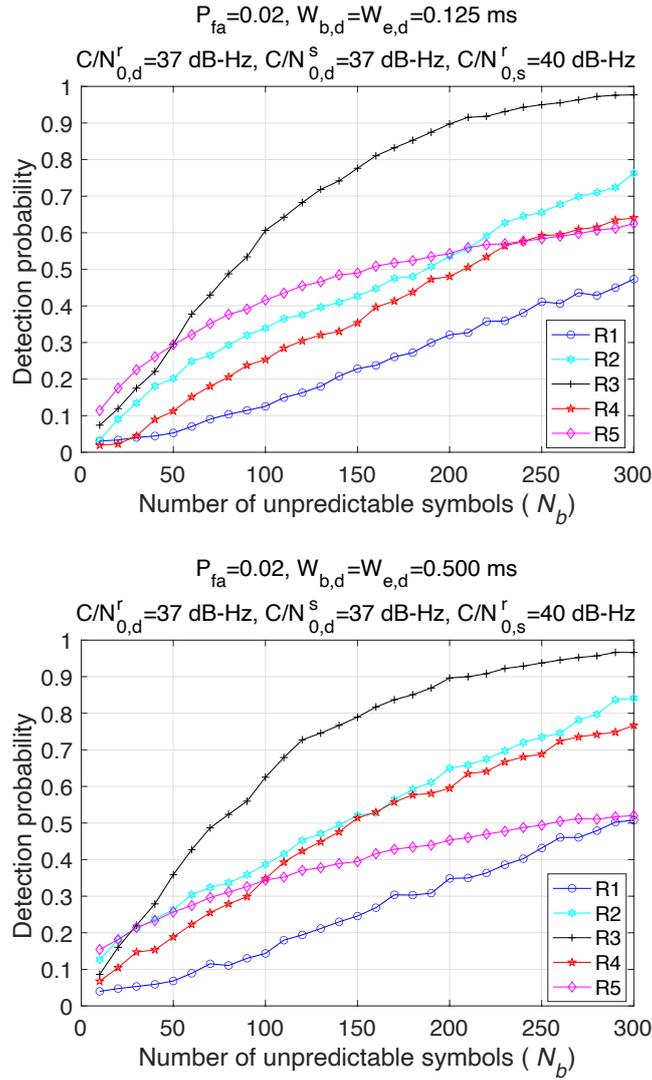

**Fig. 6** Detection probability vs. the number of unpredictable symbols for false alarm probability of 0.02 and for a different length of windows (125 $\mu s$ and 500 $\mu s$, in the top and bottom plots, respectively). In both cases, the spoofer has a 3 dB advantage with respect to the user.

Finally, with the goal of stressing the capabilities of the detectors, they are tested in a really advantageous situation for the spoofer. Fig. 7 shows the results obtained when the spoofer has a 5 dB advantage over the user in the reception of the real signal, and it transmits the spoofed signal at the same power as the real signal, which is the most deleterious condition as seen above. The results show that, while the number of symbols that need to be accumulated is higher, up to around 380 symbols for a 90% detection probability, the detector can still work under these highly advantageous conditions for the spoofer (disadvantageous for the user). The performance is worse if the integration windows, $W_{b,d}$ and $W_{e,d}$, are 250 ms or 500 ms because when the spoofer receives the authentic signal with high power, it needs a short interval of time to estimate the unpredictable symbol reliably. Ideally, the length of the observation window and the length of the interval should be similar, where the spoofer makes a large number of errors in the estimation of the sign of the unpredictable symbol. For a shorter window, part of the signal that could be used for spoofing detection is not employed. On the other hand, for a longer window, more noise than what is strictly necessary enters the detectors. Hence, for maintaining the probability of false alarm, the detection threshold must be increased, which impacts the probability of detection. Nevertheless, we have seen above that R3 is the less sensitive detector to misadjustment of the window length.

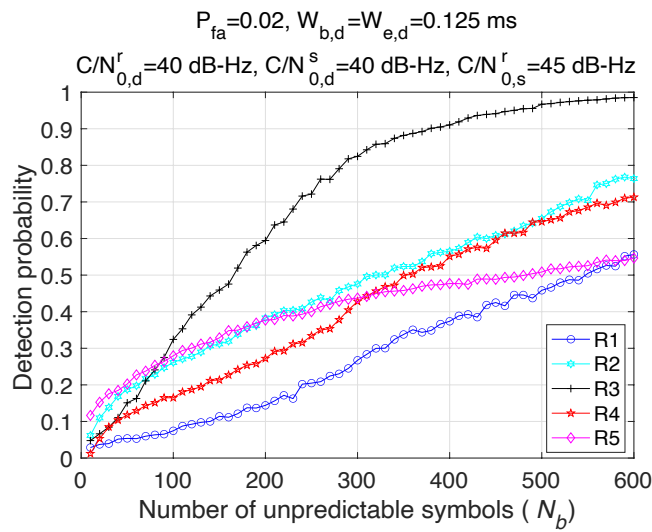

**Fig. 7** Detection probability vs the number of unpredictable symbols for false alarm probability of 0.02 with a 5 dB advantage for the spoofer

The conclusion drawn from the simulation analysis is that the R3 detector is the best performing one out of the proposed detectors. Moreover, it is robust against the degradation of the operating conditions, since the R3 detector maintains good performance even when the spoofer has a power advantage of 5 dB over the user receiver provided that the detector processes enough unpredictable symbols.

*Simulation results in a Land Mobile Satellite (LMS) channel*

The LMS channel makes the received signal contain variations in the signal amplitude caused by close-in multipath. The goal of this section is to analyze how the performance of the detectors is affected by the channel. The LMS channel used to carry out the simulation is based on the model described in Arndt et al. (2012) and Prieto-Cerdeira et al. (2010), which consists of two states. The first one is a "good state" that corresponds to a line-of-sight situation, whereas the second one is a "bad state" that considers heavy shadowing and blockage. Before proceeding, we analyze the coherence time, which is approximately the inverse of the Doppler spread of reflections, and a statistical measure of the time interval during which the channel is essentially invariant. It is given by:

$$T_c \approx \frac{c}{v\, f_c} \qquad (17)$$

where $c$ is the speed of the light, $f_c$ is the center frequency of the emitter, and $v$ is the speed of the receiver with respect to its environment. For instance, considering the speed of the receiver of 100 km/h, the coherence time is approximately 7 ms. This fact seems to indicate that speeds of about 100 km/h should not affect much the performance of the detectors because the distance between the beginning of the first part of the unpredictable symbol and the beginning of the last part of this symbol is less than 4 ms. In order to prove that, we perform the same simulation in an AWGN channel (Fig. 8-top) and in an LMS channel, also including the Gaussian noise, for a speed of 100 km/h (Fig. 8-bottom). The result confirms that the performance of the analyzed detectors, and especially R3, are not significantly degraded in multipath propagation conditions with the velocities found in terrestrial applications. It is only for velocities higher than 300 km/h, when the coherence time is shorter than the symbol time, that the performance of R3 noticeably degrades.

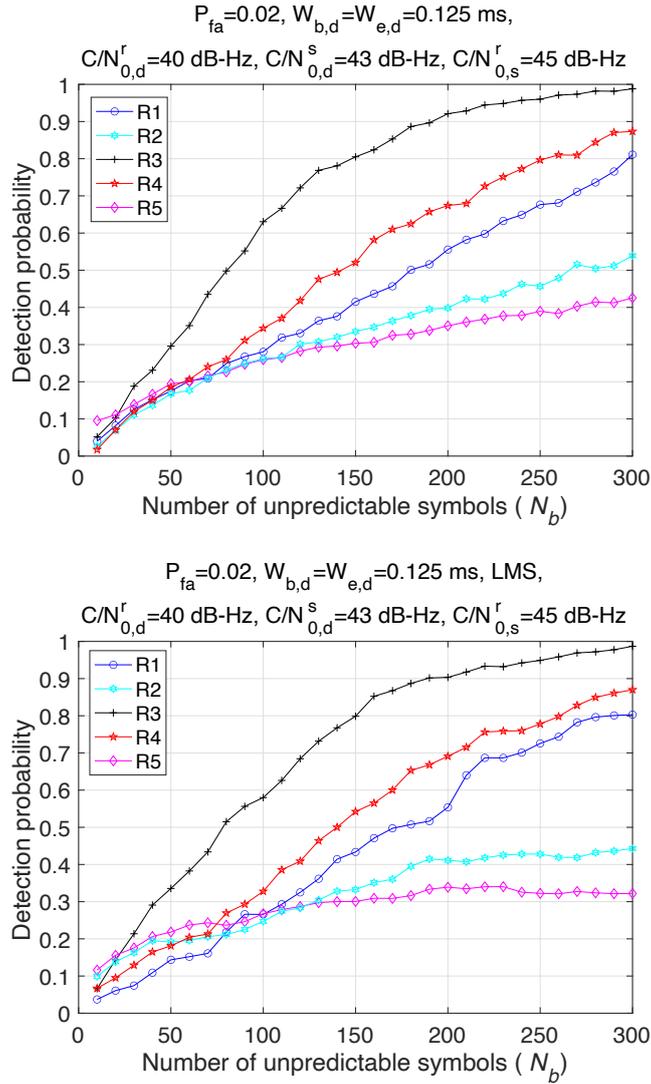

**Fig. 8** Detection probability vs the number of unpredictable symbols for false alarm probability of 0.02 for a Gaussian channel (top) and for an LMS channel with speed of 100 km/h (bottom)

In order to synthesize the previous results and add some more cases, we show in Table 3 the approximate number of symbols that are necessary for the R3 detector to reach a 0.9 probability of detecting the spoofing attack while keeping the probability of alarm at 0.02, for different combinations of the signal powers and window lengths at the detector. If the inter and intra-system interference is considered, the number of required symbols changes at most in 5 symbols.

Table 3 Number of symbols needed to detect the attack in different representative cases

| $C/N_{0_d}^r$ [dB-Hz] | $C/N_{0_d}^s$ [dB-Hz] | $C/N_{0_s}^r$ [dB-Hz] | Channel | $W_{b,d}$, $W_{e,d}$ [µs] | Required number of symbols |
|---|---|---|---|---|---|
| 37 | 37 | 40 | AWGN | 125 | 205 |
| 37 | 37 | 40 | AWGN | 250 | 200 |
| 37 | 37 | 40 | AWGN | 500 | 200 |
| 37 | 40 | 40 | AWGN | 125 | 100 |
| 37 | 40 | 40 | AWGN | 250 | 70 |
| 40 | 40 | 40 | AWGN | 125 | 105 |
| 40 | 40 | 40 | AWGN | 250 | 110 |
| 40 | 40 | 43 | AWGN | 125 | 210 |
| 40 | 40 | 43 | AWGN | 250 | 670 |
| 40 | 43 | 45 | AWGN | 125 | 190 |
| 40 | 43 | 45 | LMS | 125 | 190 |
| 40 | 40 | 45 | AWGN | 125 | 380 |

**Practical implementation in a GNSS receiver**

In this section, we look at the practical implementation and constraints of the analyzed detectors in a GNSS receiver. In particular, we address setting the detection threshold for the spoofing detector, and how long it takes the user to obtain enough unpredictable symbols from the current Galileo OSNMA protocol (Fernández-Hernández et al. 2016). We also discuss how long it takes the spoofer to sufficiently delay the spoofed signal with respect to the authentic signal without being noticed by the receiver clock in order to have a reliable estimation of the symbol.

*Spoofing detection threshold*

As we have seen that R3 is the most promising detector, in the following, we focus on the computation of its detection threshold $\gamma$. The spoofer detection boils down to the comparison of the metric R3 with a detection threshold to distinguish whether the user receiver is being spoofed or not. The detection threshold determines the probability of false alarm as:

$$P_{fa} = 1 - cdf_{R_3}(\gamma|H_0) \qquad (18)$$

where $cdf_{R_3}(\gamma|H_0)$ is the cumulative density function (cdf) of R3 in the absence of spoofing (hypothesis $H_0$). Therefore, the computation of the detection threshold given a target probability of false alarm requires the knowledge of the cdf of R3 under $H_0$:

$$\gamma = cdf_{R_3}^{-1}(1 - P_{fa}|H_0) \tag{19}$$

Under $H_0$, the distribution of the R3 metric is very similar to the Rayleigh distribution. This occurs because the value of the partial correlations at the beginning and the end of each symbol (or another predictable part of the signal) have practically the same constant value plus Gaussian noise. Then, the term inside the absolute value in (8) can be considered as a zero-mean complex Gaussian variable, and hence the metric of R3 has a Rayleigh distribution. Exploiting the relation between the Rayleigh distribution and the underlying Gaussian variable, the mean of the Rayleigh distribution can be easily obtained from the standard deviation of the partial correlations in the predictable part $B_{end}(k)$. That is, the mean of the Rayleigh distribution is equal to $\sigma_B\sqrt{\pi/2}$, where $\sigma_B$ is the variance $B_{end}(k)$. It is worth mentioning that due to the cross-correlation between the E1B and E1C components, the term inside the absolute value can have small mean value, but as we will see below, its effect is negligible.

Fig. 9 compares the simulated and the theoretically adjusted probability density functions (pdf's) under $H_0$, and $P_{fa}$ of the R3 detector with the thresholds obtained via simulation and theoretically with the Rayleigh distribution. The results show a perfect match between both, conforming that the detection threshold for the R3 detector can be fixed in an easy way.

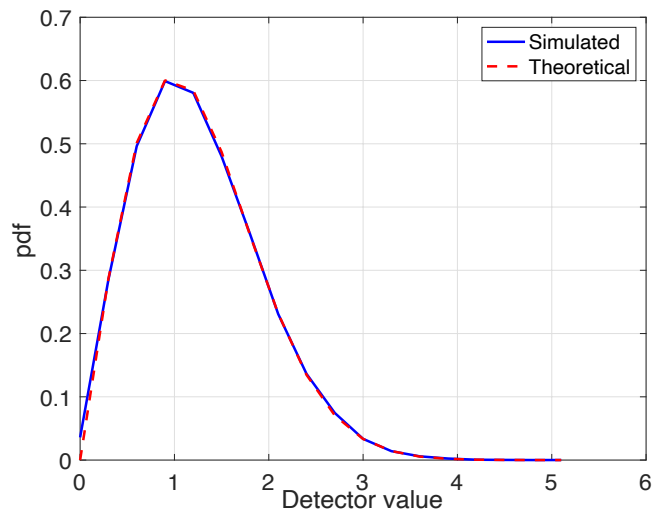

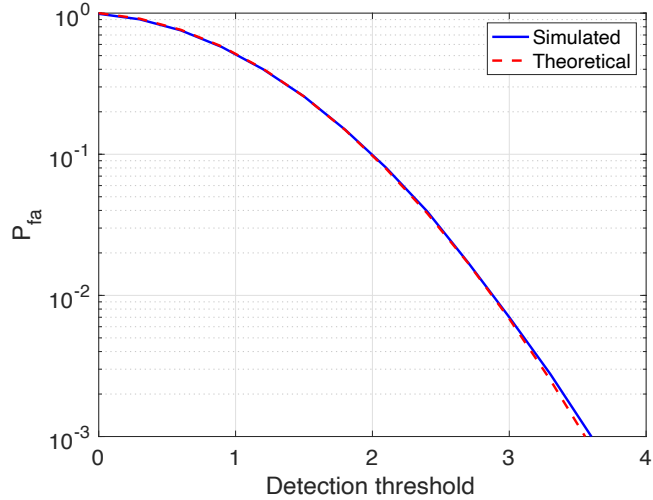

**Fig. 9** Comparison between the empirical pdf and probability of false alarm obtained from Monte Carlo simulations and the theoretical ones using the Rayleigh distribution

*Randomization of the correlations*

A spoofer knowing beforehand which unpredictable symbols, and which parts of them, will be used in the partial correlations, could exploit an advantage over the user. First, because it could implement a random-value attack with variable power, depending on the success or failure of the previous guess. Second, because it could alter the predictable correlations to spoof the detector. The randomization of the correlations can mitigate both advantages. However, this adds a slight additional complexity to the receiver and may require more unpredictable symbols, as some would be discarded. In light of the results presented in the previous section, the loss of some symbols to implement the randomization is affordable.

*Duration of anti-replay protection vs. clock stability*

This sub-section analyzes the time needed by the spoofer to sufficiently delay the spoofed signal with respect to the authentic signal in order to estimate the unpredictable symbols, without being detected by a user receiver that has the capability of detecting changes in the clock offset of the positioning solution. The capability of the user to detect these changes depends on the stability of the receiver clock.

By delaying the signal, the spoofer would obtain some advantage over the target receiver making the detection of the spoofing attack much more challenging or even impossible. To do so, first, we must know the observation window needed by the spoofer to make a reliable decision about the value of the unpredictable symbol. The duration of this window depends on the carrier-to-noise ratio of the received signal at the spoofer. We consider a very optimistic case for the spoofer, where it receives the signal with a $C/N_{0_S}^r$ =45 dB-Hz and it accepts a probability of symbol error ($P_e$) of 0.1. Assuming a Gaussian channel and neglecting cross-correlation effects of satellites, the required time ($T_{spof}$) to estimate the unpredictable symbol can be computed using the expression of the probability of error of non-coded BPSK (given that we are working at symbol level):

$$T_{spof} = \frac{\left(erfc^{-1}(2P_e)\right)^2}{C/N_{0_S}^r} = 25.97 \text{ μs} \approx 26 \text{ μs} \qquad (20)$$

The spoofer needs to delay the spoofed signal by 26 μs with respect to the authentic signal at the target receiver to have a good estimation of the unpredictable symbols. The spoofer cannot transmit a signal directly with a delay of 26 μs because the clock offset jump will be very easily detected by comparing it with the target receiver clock. In order to circumvent this problem, the spoofer can obstruct true signal reception by jamming the user receiver, without jamming its own receiver and thus preventing the reception of signals. Also, the spoofer can block the antenna for a period that depends on the clock stability, before starting the spoofing attack (non-zero-delay attack), or it must gradually introduce an increasing delay, slowly enough not to be detected by the clock. In both cases, the time that the spoofer requires to take control of the target receiver depends on the assumed stability of the receiver clock, as we analyze below.

Common GNSS receivers generally use either one of two kinds of clocks: TCXO or OCXO (thermal-controlled or oven-controlled crystal oscillators). OCXO clocks are usually more stable than TCXO clocks. The short-term stabilities of a typical TCXO and OCXO in a GNSS receiver are approximately $10^{-7}$ and $10^{-10}$ over 1 second, respectively. These values of stability are obtained approximately for averaging times around 1 second (Beard and Senior 2017). Then, considering the best case for the spoofer, i.e., that the target receiver is using a TCXO with the stability of $10^{-7}$, the spoofer would need around 260 seconds to delay the spoofing signal by 26 μs. The spoofer would, therefore, need to jam or obstruct the GNSS signals to the receiver so that the user does not get a proper time measurement during this 260-second interval to succeed.

We consider another example more pessimistic for the spoofer. We assume that the spoofer receives the signal with a $C/N_{0_s}^r$=40 dB-Hz and the time required to estimate the unpredictable symbol for a low probability of symbol error (0.01) is given by

$$T_{spof} = \frac{\left(\text{erfc}^{-1}(0.02)\right)^2}{10^4} \approx 271 \text{ μs} \qquad (21)$$

In this case, considering that the target receiver has a TCXO clock, the spoofer would need around 2710 seconds to delay the spoofing signal by 271 μs.

It is difficult for the spoofer to introduce a constant drift in the signal while the detector is running at the receiver. The reason is that detector will succeed in detecting the signal distortions before the spoofer includes a sufficient delay, provided that the detector can accumulate a sufficient number of unpredictable symbols in a period shorter than 260 or 2710 seconds, depending on the case under consideration. In favor of the spoofer, one could argue that the success probability in estimating the symbol increases during the drifting period. However, as it is shown in the next section with a real implementation based on Galileo OSNMA, the signal contains the required number of unpredictable symbols in interval durations far shorter than the bounds derived from the clock stability.

From these results, we can conclude that if the spoofer intends to perform a signal replay attack on a receiver equipped with a TCXO and already tracking real signals, the process during which the spoofer gains the necessary time margin to estimate the unpredictable symbols must last for more than four minutes and, in some cases, for more than 45 minutes, depending on the received signal power. The time intervals increase by around 3 orders of magnitude if the user employs an OCXO. This analysis, in combination with the rest of the sections, suggests that good GNSS receiver clock stability highly increases receiver resilience.

*Implementation based on Galileo E1 OSNMA*

One key remaining aspects of the Galileo E1 OSNMA protocol is how relevant symbol unpredictability is in its design. The OSNMA protocol is a specific implementation of the Timed Efficient Stream Loss-tolerant Authentication (TESLA) method (Perrig et al. 2016). There is a pending design decision on whether the TESLA key should be different form each satellite, and therefore unpredictable, at the expense of a higher CPU load, or whether each satellite should use the same key (Cancela et al. 2019). This decision choice has an impact on the rate of generation of unpredictable symbols, and on how a receiver can exploit the unpredictability and which constraints the unpredictability imposes on spoofers attempting zero-delay replay attacks. The current OSNMA protocol aims at authenticating the satellite navigation data and includes several sections. The unpredictable bits of the OSNMA are included in the Message Authentication Code and Key (MACK) section, which refers to the set of bits of the protocol that contain the message authentication codes (MACs) and the time-delayed keys that together authenticate the navigation data. The MACK section uses 480 bits each 30 seconds of a total of 600 bits reserved for OSNMA. The MACK section can be divided into several MACK blocks, and each block contains unpredictable bits in the MAC fields and possibly also in the KEY fields. A more detailed definition of the protocol can be found in Fernández-Hernández et al. (2016). We have considered a use case of OSNMA of 2 MACK blocks, 20-bit MACs, 96-bit keys, and 4 MACs per block. This configuration allows the receiver to have 80 unpredictable symbols every 15 seconds (i.e., for every MACK block), without taking into account the bits of KEY field. We can conclude that when the key is conservatively considered as predictable, the detector can obtain 160 or 240 unpredictable symbols in a signal fragment of 30 or 45 seconds (i.e., 2 or 3 MACK blocks). A receiver could wait for two Galileo I/NAV subframes, or 60 seconds in total, providing 320 unpredictable symbols, in order to increase confidence in the detector. Alternatively, we can consider that the first 64 bits of the key are unpredictable because the last 32 bits of the key are predictable because the spoofer could deduce them by a brute-force search, which is in line with current processors nowadays (Neish et al. 2019). In this case, there are 144 unpredictable every 15 seconds, or equivalently 576 unpredictable every 60 seconds.

The spoofing detection capabilities of the R3 techniques as a function of the observation time of the signal are summarized in Fig. 10. Two cases, 0 dB and 3 dB, are considered for the advantage in received signal power at the spoofer with respect to the user. Also, two different assumptions regarding the unpredictability of the key bits are made. Among the four possible combinations, in the most unfavorable one for the user, a probability of detection very close to 1 is achieved after receiving 60 seconds of signal, while only 15 seconds are needed in a less stringent situation. Note also that the closest satellite, that is, the one closest to the zenith, will provide a key that can be considered unpredictable in all implementations, which goes in detriment of the spoofer.

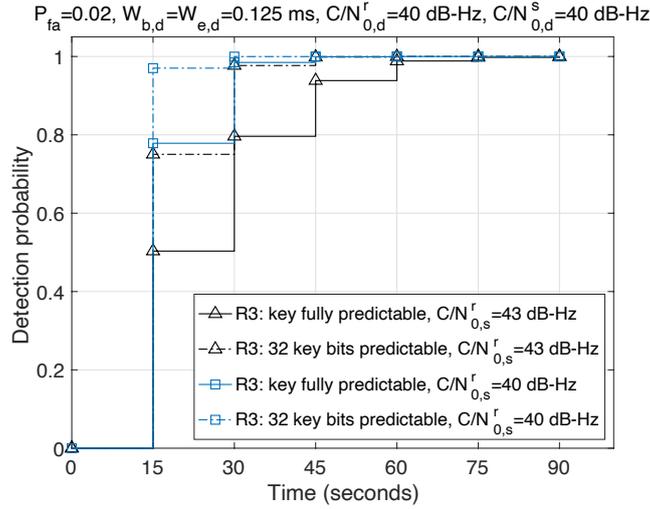

**Fig. 10** Probability of detection the spoofing attack with the R3 methods for different qualities of the signal received by the spoofer and different assumptions about the unpredictability of the key

A key result is that the interval of time needed by the user to accumulate enough unpredictable symbols to detected the replay attack (i.e., the 15 or 60 seconds mentioned in the preceding paragraph) is significantly shorter than the time that the spoofer would require to disguise the replay attack (i.e., the 260 or 2710 seconds obtained in the previous subsection). This is the confirmation of two important facts, namely: the utility of the unpredictability of symbols of the signal as an enabler for spoofing detection, and that Galileo OSNMA provides enough unpredictability even in situations favorable to the spoofer.

**Conclusions**

We have analyzed the effectiveness of the symbol unpredictability potentially present in the GNSS message as a protection against replay spoofing attacks. The analysis has focused on zero-delay attacks on a receiver tracking true GNSS signals, whereby at the beginning of the attack, the spoofer generates a signal that is fully aligned in time and frequency with the true signal, in order to take control of the tracking loops. We have proposed five spoofing detectors based on the comparison of partial correlations at the unpredictable and predictable parts of the signal, where the former are taken at the beginning of symbols that the spoofer cannot predict. Each partial correlation spans time intervals of 125, 250 and 500 μs, and they are accumulated during several unpredictable symbols.

We have analyzed the performance of the detectors under different simulated spoofing test cases, representative of the attacks proposed in the literature and of the physical limitations of a spoofer, including AWGN and LMS channels, for a GNSS receiver that is tracking Galileo E1-B I/NAV signals. We considered cases in which the spoofer has an advantage of up to 5 dB with respect to the user receiver in the reception of the signals. The tests have evaluated the detection probability versus the number of symbols accumulated for a given false alarm probability. The results have shown that the detector based on the difference, rather than the ratio or other combinations, between the initial and final partial correlations outperforms all the rest best in most situations. We have observed that the detection probability improves if the time span of the partial correlations is adapted to the $C/N_0$ of the real signal as received by the spoofer, but in general, the duration of 125μs is the preferred one. Nevertheless, another advantage of the aforementioned detector is that it is more insensitive than the others to the choice of that time span.

We have shown the number of unpredictable symbols needed by each detector in different cases. The finally selected detector requires a number that ranges from around 380 symbols in very favorable conditions to the spoofer to around 70 symbols when the spoofer is on an equal footing with the user.

Practical aspects of implementing the selected have been studied. These include setting the detection threshold, randomizing the correlations to increase robustness, clock stability considerations limiting the capability of the spoofer to shift the signal without being noticed, and implementation using the Galileo OSNMA message. OSNMA is expected to provide around 240 unpredictable symbols every 45 seconds, even if the cryptographic keys are considered predictable, and 240 unpredictable symbols are sufficient to implement a very powerful detector. Even if the user receiver has a low-quality clock, the spoofer needs a time interval significantly longer than 45 seconds to introduce a delay large enough to estimate the unpredictable symbols and to stop introducing a distortion at the beginning of those symbols. As the time needed by the spoofer to delay the signal is larger than the time needed by the proposed spoofing detector to operate reliably, the attack becomes detectable.

Therefore, we can conclude that, for a receiver already tracking the true signals, signal antireplay detection based on symbol unpredictability is a viable strategy against replay attacks. This is true as a standalone mechanism, and also as a mechanism that can be combined with other protection strategies, such as the measurements or PVT consistency checks, and the combination with other sensors.

Further work to strengthen receiver resilience against replay attacks can follow several directions: a deeper analysis of the randomization of the partial correlations; study of sensitivity of the detectors to the inclusion of both unpredictable and predictable symbols leading partial correlations; tuning of the spoofing detectors based on the experimentation with real signals or existing recorded spoofing test cases, such as ds8 in Humphreys (2015); a more exhaustive analysis of replay detectors in the presence of severe multipath and with lower false alarm probabilities; refinement of carrier phase detectors, like R5, accounting for cycle slips.

**Acknowledgments**

This work was supported in part by the European GNSS Agency (GSA) framework contract GSA/OP/12/26/Lot.1, and in part by the Research and Development Projects of Spanish Ministry of Science, Innovation, and Universities under Grants TEC2017-89925-R and TEC2017-90808-REDT, and by the ICREA Academia Program.

**Author Biographies**

**Gonzalo Seco-Granados** received the Ph.D. degree in Telecommunications Engineering from the Universitat Politècnica de Catalunya, in 2000, and the MBA degree from IESE Business School, in 2002. Until 2005, he was with the European Space Agency, involved in the design of the Galileo system. Since 2006, he is with Universitat Autònoma de Barcelona and also affiliated with the Institute of Space Studies of Catalonia.

**José A. López-Salcedo** received his M.Sc. and Ph.D. degrees in Telecommunications Engineering from the Universitat Politècnica de Catalunya, in 2001 and 2007, respectively. Since 2006, he is with Universitat Autònoma de Barcelona, and he also affiliated with the Institute of Space Studies of Catalonia, and he held a visiting appointment at the EC Joint Research Center.

**David Gómez-Casco** received the M.Sc. and Ph.D. degrees in telecommunication engineering, in 2014 and 2018, respectively, from the Universitat Autònoma de Barcelona (UAB). From July 2015 to January 2016, he carried out a research stay at the European Space Agency (ESA), The Netherlands. Since 2019, he has been a GNSS engineer with Indra.

**Ignacio Fernández-Hernández** has led over the last years the definition and implementation of authentication and high accuracy services for Galileo. He currently works for the European Commission, DG GROW. He has an MSC from ICAI, Madrid, and a Ph.D. from Aalborg University, both in Electronics Engineering, and an MBA by LBS, London.